\documentstyle[floats,preprint,aps,epsf]{revtex}
\tightenlines

\newcommand{\beq}{\begin{equation}}
\newcommand{\eeq}{\end{equation}}
\newcommand{\beqa}{\begin{eqnarray}}
\newcommand{\eeqa}{\end{eqnarray}}
\newcommand{\sr}{\sqrt}
\newcommand{\fr}{\frac}

\makeatletter
\@ifundefined{lesssim}{}{}
\@ifundefined{gtrsim}{}{}
\def\vereq#1#2{\lower3pt\vbox{\baselineskip1.5pt \lineskip1.5pt
\ialign{$\m@th#1\hfill##\hfil$\crcr#2\crcr\sim\crcr}}}
\makeatother

\begin{document}

\draft \preprint{hep-th/0403015, KIAS-P04016}

\title{%
Classical Stability of Black Branes\footnote{Based on a talk
presented at the 6th International Conference on Gravitation and
Astrophysics held at Ewha Womans University, Seoul in Korea during
Oct. 6-9, 2003.} }

\author{%
Gungwon Kang 
}
\address{%
School of Physics, Korea Institute for Advanced Study, \\
207-43 Cheongryangri-dong, Dongdaemun-gu, Seoul 130-012, Korea
}

\maketitle

\begin{abstract}

Classical stability behaviors of various static black brane
backgrounds under small perturbations have been summarized
briefly. They include cases of black strings in AdS$_5$ space,
charged black $p$-brane solutions in the type II supergravity, and
the BTZ black string in four-dimensions. The relationship between
dynamical stability and local thermodynamic stability - the
so-called Gubser-Mitra conjecture - has also been checked for
those cases.

\end{abstract}

\section{Introduction}

In the four-dimensional spacetime, event horizons of non-spherical
topology are forbidden for asymptotically flat stationary black
holes. In higher dimensions, however, the topology could be either
hyperspherical ($S^{D-2}$) or hypercylindrical ($S^n \times
R^{D-2-n}$ or $S^n \times S^{D-2-n}$). The four-dimensional
Schwarzschild black hole in Einstein gravity is known to be stable
classically under linearized perturbations. Recently, Ishibashi
and Kodama~\cite{Ishibashi:2003ap} have shown that this stable
behavior extends to hold for higher dimensional cases as well.
Black hole solutions in higher dimensions having hypercylindrical
horizons are called black strings or branes. Gregory and
Laflamme~\cite{GL1} have investigated the stability of a black
$p$-brane that is a product of the $(D-p)$-dimensional
Schwarzschild black hole with the $p$-dimensional flat space, and
found that such background is unstable as the compactification
scale of extended directions becomes larger than the order of the
horizon radius - the so-called Gregory-Laflamme instability.
Gregory and Laflamme~\cite{Gregory:1994bj} also considered a class
of magnetically charged black $p$-brane solutions for a stringy
action containing the NS5-brane of the type II supergravity. For
horizons with infinite extent, they have shown that the
instability persists to appear but decreases as the charge
increases to the extremal value. On the other hand, branes with
extremal charge turned out to be stable~\cite{Gregory:1994tw}.
Since their discovery of such linearized instability, black
strings or branes have been believed to be generically unstable
classically under small perturbations except for the cases of
extremal or suitably compactified ones, and the Gregory-Laflamme
instability has been used to understand physical behaviors of
various systems involving black brane configurations as in string
theory.

In the context of string theory, however, black branes that
Gregory and Laflamme considered are those having magnetic charges
with respect to Neveu-Schwarz gauge fields only. Recently a wider
class of black string or brane backgrounds has been studied in
order to see whether or not the stability behavior drastically
changes. In this talk I briefly summarize some of interesting
results obtained so far, and report some new results for black
$D3$-branes.

\section{Linearized stability behaviors}

In order to check the classical stability of a given black string
or brane background under small perturbations, we seek any
unstable linearized solution that grows in time and is regular
spatially outside the event horizon. In the viewpoint of the
Kaluza-Klein (KK) dimensional reduction, such unstable solution
can be expanded in terms of KK modes along the extra directions
characterized by the KK mass parameter $m$. In particular,
$s$-wave perturbations that are spherically symmetric in the
submanifold perpendicular to the spatial worldvolume are believed
to be the strongest instability for most cases. The existence of
such $s$-wave unstable mode can be checked by finding the
so-called threshold mode that is static and the onset of
instability. The number of unknown functions in the analysis can
be further reduced by suitably choosing gauge conditions allowed
in the system. If the set of coupled linearized equations allows
any threshold mode solution with a non-vanishing threshold KK mass
$m^*$ for certain parameter values characterizing the background
fields, the corresponding black brane is unstable.

\subsection{Black strings in AdS space}

\begin{figure}[tbp]
 \centerline{
  \epsfysize=60mm\epsfxsize=80mm\epsffile{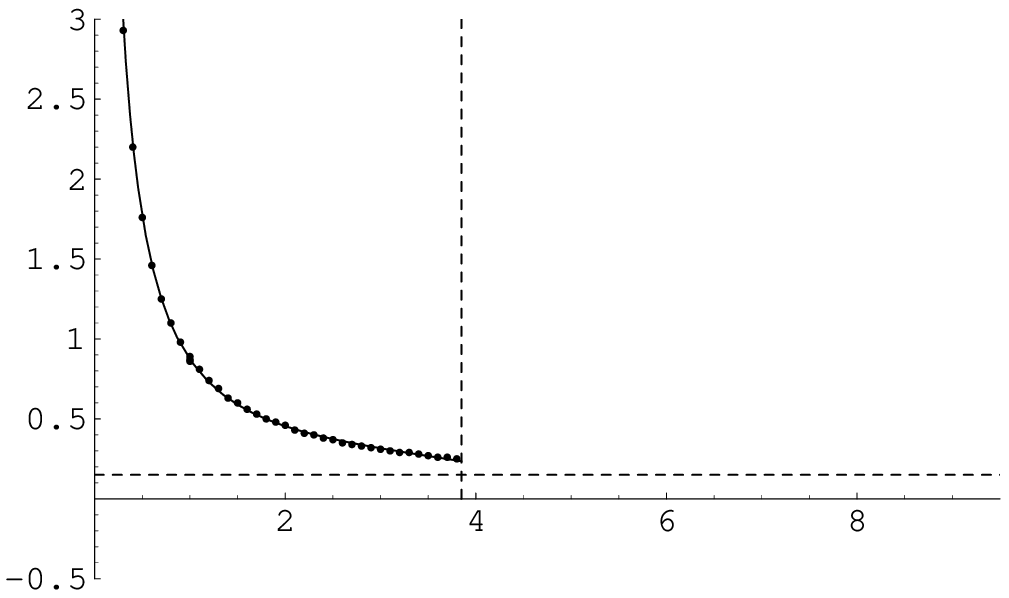}
  \epsfysize=60mm\epsfxsize=80mm\epsffile{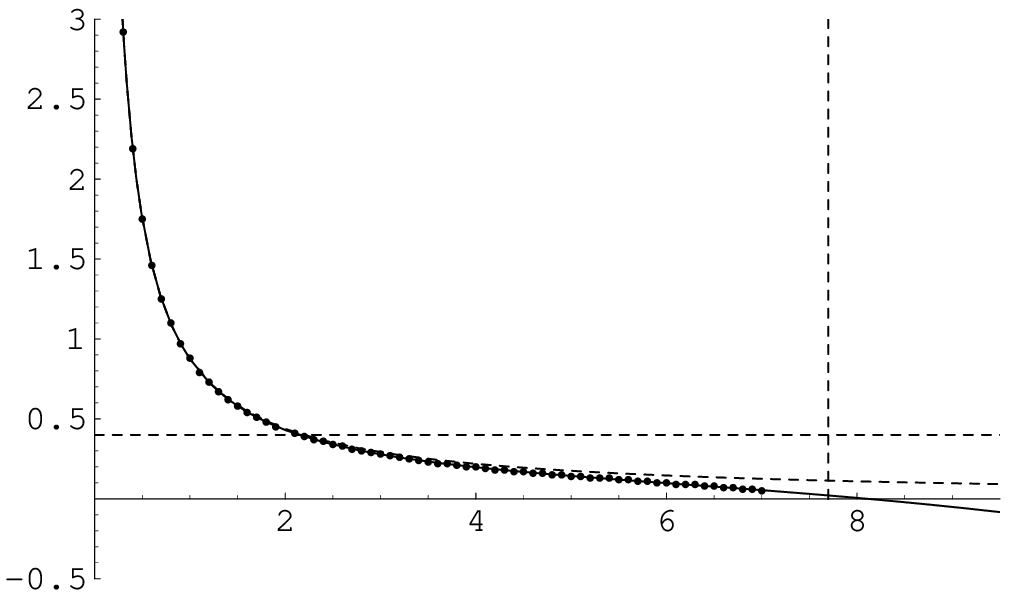}
   \put(-380,180){$dS\,(\Lambda_4 > 0)$}
   \put(-450,180){$m^{*}$}
   \put(-245,26){$r_0$}
   \put(-340,38){\small{$m_{min.}$}}
   \put(-165,180){$AdS\,(\Lambda_4<0)$\,/\,\,$\Lambda_4=0$}
   \put(-220,180){$m^{*}$}
   \put(-12,26){$r_0$}
   \put(-140,49){\small{$m_{min.}$}}
 }
\vspace{0.3cm}
 \caption{%
 The lefthand side: threshold masses for varying $r_0$ with given
 $l_4=10$ in the dS$_4$-Schwarzschild black strings. The vertical dotted
 line denotes the maximum possible value (Nariai limit), $r_0 \simeq 3.85$.
 Note that all threshold masses are larger than the lowest value in the KK
 mass spectrum ({\it i.e.}, $m^*(r_0\simeq 3.85)=0.29 > m_{\rm min.}
 \simeq 0.15$). %
 The righthand side: For the case of Schwarzschild black strings threshold
 masses (dotted curve) never cross the horizontal axis ($m_{\rm min}=0$).
 For AdS$_4$-Schwarzschild black strings, however, the threshold mass becomes
 smaller than the lowest allowed KK mass at a certain value of
 $r_0 (\simeq 2.1)$, implying no unstable solution for $r_0$ larger than such
 critical value. }
 \label{BSinAdS}
\end{figure}

In five-dimensional AdS space with or without a uniform $3$-brane,
one may have three types of static black string solutions
characterized by a four-dimensional cosmological constant
$\Lambda_4 = \pm 3/l^2_4$ and a parameter $r_0$ that is related to
the mass density. They are Schwarzschild ($\Lambda_4 =0$),
dS$_4$-Schwarzschild ($\Lambda_4
>0$), and AdS$_4$-Schwarzschild ($\Lambda_4<0$) black strings.
As shown in Fig.~\ref{BSinAdS} for varying $r_0$ with a given
value of $\Lambda_4$~\cite{Hirayama:2001bi}, there always exist
non-vanishing threshold masses for cases of
Schwarzschild~\cite{Gregory:2000gf} and dS$_4$-Schwarzschild black
strings, implying instability as usual. For the case of
AdS$_4$-Schwarzschild black strings, however, the threshold mass
becomes smaller than the lowest allowed KK mass ($m_{\rm min}$) at
a certain value of $r_0 (\simeq 2.1)$. Without a $3$-brane $m_{\rm
min} = 4/l_4 =0.4$ and this critical value of $r_0$ corresponds to
the horizon radius of $r_+^{\rm cr} \simeq 0.20 l_4$. Thus, there
is indeed no threshold mode solution for $r_+ \geq r_+^{\rm cr}$.
In other words, the $s$-wave instability present in small size
AdS$_4$-Schwarzschild black strings disappears as the horizon
radius becomes larger than the order of the AdS$_4$ radius.

\subsection{Black branes in type II string theory}

\begin{figure}[tbp]
 \centerline{
  \epsfysize=60mm\epsfxsize=80mm\epsffile{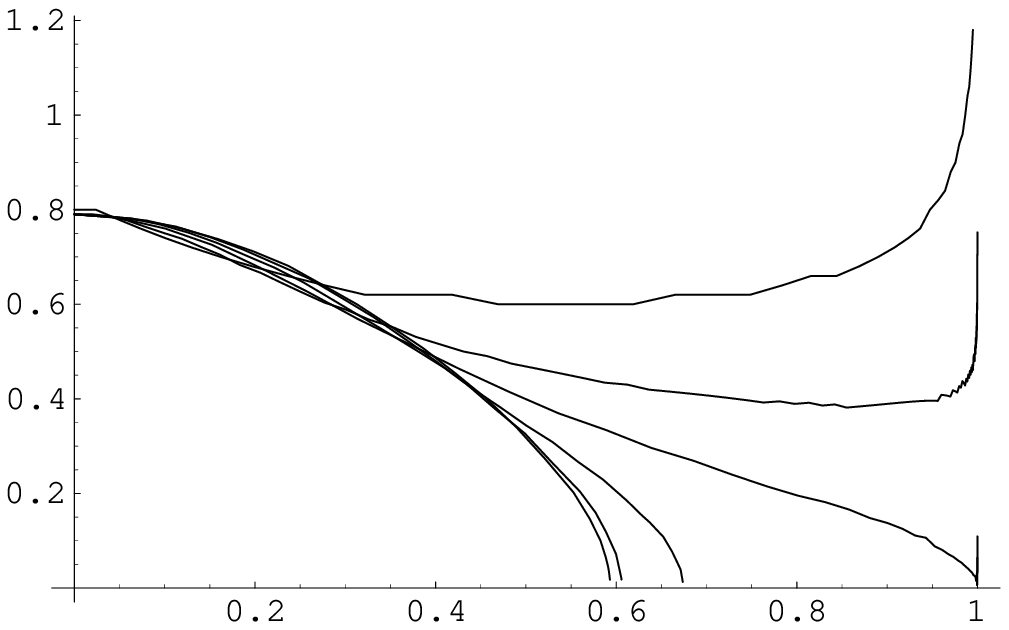}
  \epsfysize=60mm\epsfxsize=80mm\epsffile{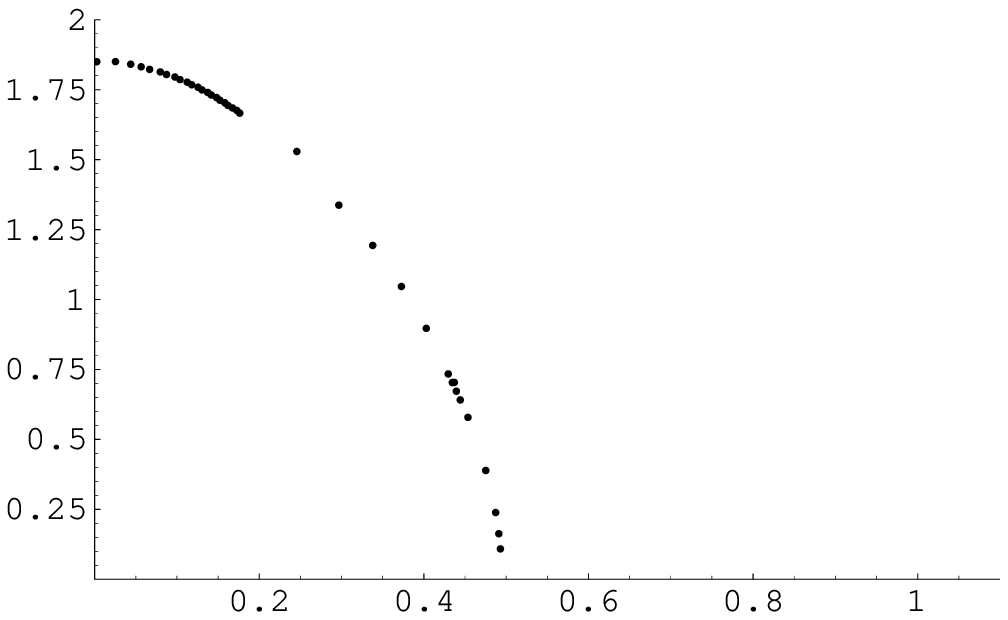}
   \put(-350,185){${\bf p=4}$}
   \put(-450,180){$m^*$}
   \put(-233,10){$q$}
   \put(-448,5){$\scriptstyle 0$}
   \put(-350,94){$\scriptstyle a=3$}
   \put(-337,75){$\scriptstyle a=2$}
   \put(-325,58){$\scriptstyle a=3/2\, ({\rm GL})$}
   \put(-317,40){$\scriptstyle a=1$}
   \put(-317,24){$\scriptstyle a=1/2\, ({\rm D}4)$}
   \put(-336,14){$\scriptstyle a=0$}
   \put(-120,185){${\bf D3}$}
   \put(-220,180){$m^*$}
   \put(-2,10){$q$}
   \put(-216,5){$\scriptstyle 0$}
    }
\vspace{0.3cm}
\caption{%
The lefthand side: behavior of threshold masses for black
$4$-branes in $D=10$ at various values of $a$ with fixed mass
density $M=2^5$. $m^* \simeq 1.581/r_{\scriptscriptstyle{H}}\simeq
0.791$ with $r_{\scriptscriptstyle{H}}=2$ for uncharged black
branes ({\it i.e.}, $q=0$). The extremality parameter $q=1$ for
extremal branes. Critical values at which the instability
disappears are $q_{\rm cr} \simeq 0.593$, $0.606$, $0.674$ for
cases of $a=0$,
$1/2$, $1$, respectively. %
The righthand side: behavior of threshold masses for black
$D3$-branes with mass density $M=5$. The critical value is $q_{\rm
cr} \simeq 0.49$. }
 \label{Dpbranes}
\end{figure}

Recently, Hirayama, Kang, and Lee~\cite{Hirayama:2002hn} have
analyzed the linearized stability of a wider class of magnetically
charged black $p$-brane solutions for the string gravity action
given by
 \beq
 {\rm I} = \int d^Dx\sr{-g} \left[ R -\fr{1}{2} (\partial
\phi)^2 -\fr{1}{2n!}e^{a\phi}F_n^2 \right]
 \label{action}
 \eeq
in the Einstein frame. It turns out that the stability of these
black $p$-branes behaves very differently depending on the
coupling parameter $a$. That is, there exists a critical value of
the coupling parameter $a_{\rm cr}(D,p) = (D-3-p)/\sr{(D-2)/2}$
determined by the full spacetime dimension $D$ and the dimension
of the spatial worldvolume $p=D-2-n$. The case that Gregory and
Laflamme studied is precisely when $a=a_{\rm cr}$. In this case
black branes with horizons of infinite extent are always unstable
as explained above, and magnetically charged NS5-branes of the
type II supergravity belong to this class. When $0 \leq a < a_{\rm
cr}$, black branes with small charge are unstable as usual. As the
charge increases, however, the instability decreases and
eventually disappears at a certain critical value of the charge
density which could be even far from the extremal point.
Magnetically charged black D0, F1, D1, D2, D4 branes of the type
II string theory belong to this class for instance. When $a >
a_{\rm cr}$, on the other hand, the instability persists all the
way down to the extremal point. Interestingly it should be noticed
that in this case the threshold mass starts to increase again near
the extremal point ($q \simeq 1$) as can be seen in
Fig.~\ref{Dpbranes}. Such stability behavior in the presence of
charge never has been expected in the literature. Magnetically
charged black D5 and D6 branes in the type II supergravity are in
this case for example. However it is shown that all black branes
mentioned above are stable at the extremal point, which might be
expected due to the BPS nature of extremal solutions in string
theory. On the lefthand side of Fig.~\ref{Dpbranes} behaviors of
threshold masses are illustrated for some black $4$-branes as the
charge increases.

\subsection{Black $D3$-branes}

The case of $n=D/2$ and $a=0$ with a self-dual $n$-form field
strength ${\bf F}=\mbox{}^{*}{\bf F}$ in Eq.~(\ref{action}) is
treated separately for some technical reasons. This case includes
black D3-branes ({\it i.e.}, $n=5$) in the type II supergravity
for which the AdS/CFT correspondence has been understood very well
in string theory. In contrast to previous cases mentioned above,
the fluctuation of the dilaton field is completely decoupled and
can be set to be zero, but the $s$-wave perturbation of the field
strength should not be frozen in order to be consistent with the
metric perturbation as a black brane gets charged. As shown in the
numerical results on the righthand side of Fig.~\ref{Dpbranes} for
black $D3$-branes, a black brane in this class is unstable when it
has small charge density. As the charge density increases for
given mass density, however, the instability decreases down to
zero at a certain finite value of the charge density, and then the
black brane becomes stable all the way down to the extremal
point~\cite{KL,Gubser:2002yi}.

\subsection{Black strings in $D \leq 4$}

It is possible to have stationary black string or brane solutions
even in four dimensions when a negative cosmological constant is
present. Interestingly it is likely that all known stationary
black branes in four or three dimensions are stable since they are
thermodynamically stable. In particular, the case of BTZ black
strings has been explicitly checked to be stable by linearized
analysis for any type of perturbations~\cite{Kang:2002hx}.

\section{Thermodynamic stability behavior}

One of naive explanations for the occurrence of the
Gregory-Laflamme instability is that for a given black string or
brane configuration there exists some other black hole
configuration such as a hyperspherical black hole that has the
same mass and charge, but possesses larger
entropy~\cite{GL1,Gregory:1994bj}. Recently, Gubser and
Mitra~\cite{Gubser:2000ec} refined such entropy comparison
argument, and proposed a conjecture that a black brane with a
non-compact translational symmetry is classically stable if and
only if it is locally thermodynamically stable. The proof of this
Gubser-Mitra (GM) conjecture has been argued for a certain class
of black brane systems by Reall~\cite{Reall:2001ag}. This
interesting relationship between the classical dynamical stability
and the local thermodynamic stability has been explicitly checked
to hold for various black string or brane
systems~\cite{PGR,Gubser:2000ec,Kang:2002hx,Hirayama:2002hn,Gubser:2002yi,KL}.
When the translational symmetry along the horizon is broken, one
can see some disagreements for onset points of instability as
shown in the stability analysis for AdS$_4$-Schwarzschild black
strings in AdS space~\cite{Hirayama:2001bi}. It also should be
pointed out that the GM conjecture simply gives the information
about when a black string or brane becomes stable or unstable. It
does not explain or predict other details of classical stability
behaviors~\cite{Hirayama:2002hn}.

\section{Discussion}

To conclude, it is briefly summarized how the classical linearized
stability behaves for various black string or brane solutions in
several classes of theories. In lower dimensions ({\it i.e.}, $D
\leq 4$) all black branes seem to be stable. In higher dimensions
({\it i.e.}, $D > 4$) all neutral black branes considered are
unstable unless they are suitably compactified or have some
suitable AdS nature. As a black brane gets charge, the $s$-wave
instability could either persist all the way down to the extremal
point or disappear at a certain value of charge density, depending
on its theory. For all cases we considered, even if a non-extremal
black brane possesses the $s$-wave instability at any value of the
charge density, its extremal one turns out to be free of such
instability. On the other hand, the GM conjecture turned out to
hold well for the cases of having the translation symmetry along
the spatial worldvolume.

Further work is required to have deeper understandings of these
diverse stability behaviors. Some of interesting open problems are

\begin{itemize}
\item
Physical understanding of the transition point in the parameter
space at which the stability behavior changes.
\item
The consequence of the Gregory-Laflamme instability and its
evolution afterwards when it is present.
\item
General proof of the stability when the $s$-wave instability is
absent.
\item
Possibility of some simple criterion for the stability such as the
GM conjecture in the case of stationary non-uniform black branes.

\end{itemize}


\begin{thebibliography}{99}

\bibitem{Ishibashi:2003ap}
A.~Ishibashi and H.~Kodama,
Prog.\ Theor.\ Phys.\  {\bf 110}, 901 (2003), hep-th/0305185.

\bibitem{GL1}
R. Gregory and R. Laflamme, Phys. Rev. Lett. {\bf 70}, (1993)
2837.

\bibitem{Gregory:1994bj}
R.~Gregory and R.~Laflamme,
Nucl.\ Phys.\ B {\bf 428}, 399 (1994), hep-th/9404071.

\bibitem{Gregory:1994tw}
R.~Gregory and R.~Laflamme,
Phys.\ Rev.\ D {\bf 51}, 305 (1995), hep-th/9410050.

\bibitem{Hirayama:2001bi}
T.~Hirayama and G.~Kang,
Phys.\ Rev.\ D {\bf 64}, 064010 (2001), hep-th/0104213.

\bibitem{Gregory:2000gf}
R.~Gregory,
Class.\ Quant.\ Grav.\  {\bf 17}, L125 (2000), hep-th/0004101.

\bibitem{Hirayama:2002hn}
T.~Hirayama, G.~Kang and Y.~Lee,
Phys.\ Rev.\ D {\bf 67}, 024007 (2003), hep-th/0209181.

\bibitem{KL}
G. Kang and J. Lee;``Classical Stability of Black D3-branes,''
preprint KIAS-P03062, hep-th/0401225.

\bibitem{Gubser:2002yi}
S.~S.~Gubser and A.~Ozakin,
JHEP {\bf 0305}, 010 (2003), hep-th/0301002.

\bibitem{Kang:2002hx}
G.~Kang, {\it Proceedings of the 11th Workshop on General
Relativity and Gravitation} held at Tokyo, Japan, Jan. 9-12, 2002,
arXiv:hep-th/0202147; G.~Kang and Y.~Lee;``Lower Dimensional Black
Strings/Branes Are Stable,'' preprint KIAS-P03063 (2003).

\bibitem{Gubser:2000ec}
S.~S.~Gubser and I.~Mitra;``Instability of charged black holes in
anti-de Sitter space,'' hep-th/0009126;
JHEP {\bf 0108}, 018 (2001), hep-th/0011127.

\bibitem{Reall:2001ag}
H.~S.~Reall,
Phys.\ Rev.\ D {\bf 64}, 044005 (2001), hep-th/0104071.

\bibitem{PGR}
T.~Prestidge,
Phys.\ Rev.\ D {\bf 61} (2000) 084002, hep-th/9907163;
J.~P.~Gregory and S.~F.~Ross,
Phys.\ Rev.\ D {\bf 64}, 124006 (2001), hep-th/0106220.


\end{thebibliography}
\end{document}